\documentclass[twocolumn,showpacs,preprintnumbers,amsmath,amssymb,prd,aps]{revtex4-1}
\usepackage{graphicx}
\usepackage{dcolumn}
\usepackage{bm}
\usepackage{lineno}
\usepackage{amsmath}
\usepackage{relsize}

\begin{document}

\title{Bounds on the origin of extragalactic ultrahigh energy cosmic rays from the IceCube neutrino observations}

\author{Shigeru Yoshida}
\thanks{syoshida@hepburn.s.chiba-u.ac.jp (S.~Yoshida)}
\affiliation{Department of Physics, Graduate School of Science, Chiba University, Chiba 263-8522, Japan}
\author{Hajime Takami}
\thanks{takami@post.kek.jp (H.~Takami)}
\thanks{JSPS research fellow}
\affiliation{High Energy Accelerator Research Organization (KEK), Ibaraki 305-0801, Japan}

\date{\today}

\begin{abstract}
We study general implications of the IceCube observations in the energy range from $10^{6}$~GeV to $10^{10}$~GeV for the origin of extragalactic ultrahigh energy cosmic rays assuming that high energy neutrinos are generated by the photomeson production of protons in the extragalactic universe. The PeV-energy neutrino flux observed by IceCube gives strong bounds on the photomeson-production optical depth of protons in their sources and the intensity of the proton component of extragalactic cosmic rays. The neutrino flux implies that extragalactic cosmic-ray sources should have the optical depth greater than $\sim 0.01$ and contribute to more than a few percent of the observed bulk of cosmic rays at 10 PeV. If the spectrum of cosmic rays from these extragalactic sources extends well beyond 1 EeV, the neutrino flux indicates that extragalactic cosmic rays are dominant in the observed total cosmic-ray flux at 1 EeV and above, favoring the dip transition model of cosmic rays. The cosmic-ray sources are also required to be efficient neutrino emitters with the optical depth close to unity in this case. The highest energy cosmic-ray ($\sim 10^{11}$ GeV) sources should not be strongly evolved with redshift to account for the IceCube observations, suggesting that any cosmic-ray radiation scenarios involving distant powerful astronomical objects with strong cosmological evolution are strongly disfavored. These considerations conclude that none of the known extragalactic astronomical objects can be simultaneously a source of both PeV and trans-EeV energy cosmic rays. We also discuss a possible effect of cosmic-ray propagation in magnetized intergalactic space to the connection between the observed total cosmic-ray flux and neutrino flux. 
\end{abstract}


\maketitle

\section{\label{sec:intro} Introduction}

The detection of PeV-energy neutrinos~\cite{icecubePeV2013} and the follow-up analysis~\cite{icecubeHESE2013} by the IceCube Collaboration opened high-energy neutrino astronomy. These detected neutrinos are difficult to be reconciled with cosmogenic neutrinos \cite{Roulet2013JCAP01p028} (but see also Ref. \cite{Kalashev2013PRL111p041103}), and therefore revealed the existence of astrophysical ``on-source'' neutrinos in the energy range from sub-PeV to PeV. These ultrahigh energy (UHE) neutrinos are expected to originate from the interactions of UHE cosmic ray (UHECR) protons with matter ($pp$ interactions) or photons ($\gamma p$ interactions) in their sources. Although the arrival directions of the detected UHE neutrinos are consistent with isotropic distribution and exhibit no statistically significant correlation with any particular astronomical objects discussed as possible UHECR-driven neutrino emitters~\cite{icecubeHESE2013}, the bulk intensity of the observed neutrinos, $E_\nu^2dJ_{\nu_e+\nu_\mu+\nu_\tau}/dE_\nu\sim 3.6\times 10^{-8}$~GeV cm$^{-2}$ s$^{-1}$ sr$^{-1}$, offers an important clue to understand the general characteristics of UHECR sources through the link between the observed cosmic-ray and neutrino intensities. Their comparison provides the density of interaction targets, which is important information to constrain properties with UHECR sources.

The neutrino intensity also provides an index to probe a redshift distribution of UHECR sources. Neutrino intensity averaged over the sky is a consequence of the integral of neutrino emission over cosmic history since neutrinos can penetrate over cosmological distance. It traces the cosmic evolution of a neutrino emission rate per comoving volume. This advantage is especially prominent in 100~PeV -- 10~EeV (EeV $= 10^3$ PeV) range where the ``GZK cosmogenic'' neutrinos~\cite{berezinsky69} are expected to be dominant in the neutrino sky. The cosmogenic neutrinos are generated by the Greisen-Zatsepin-Kuz'min (GZK) process, i.e., the photomeson production of UHECRs with the cosmic microwave background (CMB) photons~\cite{GZK}. As the intensity of UHECRs and the evolution of CMB over the cosmological history are well understood, a measurement of the cosmogenic neutrino intensity can constrain the source distribution function (SDF) of UHECRs in the redshift space~\cite{yoshida93,takami09,kotera2010}. The upper limit of neutrino intensity with energies beyond 100 PeV can hence be converted into the constraints on the SDF of UHECRs~\cite{uhecr_constraint2012}. The null observation of neutrinos above 100 PeV by the IceCube experiment has indeed put the most stringent bound on the cosmological evolution of the UHECR emission rate~\cite{icecubeEHE2013}.

In this work, we study implications for general characteristics of extragalactic UHECR sources obtained by neutrino observations with the IceCube observatory on the hypothesis that the PeV-energy neutrinos are produced through photomeson production ($\gamma p$) in UHECR sources. We derive an analytical formula to calculate the intensity of on-source astrophysical neutrinos. This formulation is an extension from the analytical formula to estimate a cosmogenic neutrino flux developed in Ref.~\cite{uhecr_constraint2012}. The connection between the observed neutrino flux and UHECR proton flux allows for constraining the optical depth of the photomeson collision in UHECR sources. We also bound the cosmological evolution of the UHECR sources by the measured neutrino flux at PeV energies~\cite{icecubeHESE2013} as well as by the upper limit of neutrino intensity in the energy region above 100 PeV~\cite{icecubeEHE2013}.

This paper is laid out as follows. The analytical formula for diffuse neutrino intensity is proposed in Sec. \ref{sec:function}. In Sec. \ref{sec:result}, we derive general constraints on UHECR sources from IceCube observations and their possible links to the total cosmic-ray flux. We interpret these constraints and discuss some effects to the constraints, including a propagation effect in magnetized intergalactic space, in Sec. \ref{sec:discussion}. Our findings are summarized in Sec. \ref{sec:summary}. The standard $\Lambda$CDM cosmology with $H_0 = 73.5$ km s$^{-1}$ Mpc$^{-1}$, $\Omega_{\rm M} = 0.3$, and $\Omega_{\Lambda}=0.7$~\cite{PD2010}, which are the Hubble parameter, the matter density normalized by the critical density, and cosmological constant in the unit of the critical density, respectively, is assumed throughout the paper.

\section{\label{sec:function} Analytical formula for estimating on-source $\nu$ intensity}

The intensity of neutrinos from {\it in situ} photomeson production in UHECR sources can be calculated in an analytical manner with several approximations. The analytical formulation in this study follows that for GZK cosmogenic neutrinos which has already been published~\cite{uhecr_constraint2012}. Key strategies are the approximations that the kinematics of photomeson production is treated as single-pion production and that the $\Delta$ resonance governs the interaction. A power-law distribution of target photons is adopted for on-source neutrinos, instead of the CMB spectrum for GZK cosmogenic neutrinos. The analytical formula allows us to constrain astrophysical parameters concerning UHECR emission from the observational results of IceCube without intensive computation. In this section, we describe the essence of deriving the analytical formulation and resultant equations to represent diffuse on-source neutrino intensity. The details in this formulation are found in Ref.~
 \cite{uhecr_constraint2012}.

The neutrino number intensity $dJ_{\nu}/dE_{\nu}$ can be calculated by
\begin{equation}
\frac{d J_{\nu}}{dE_{\nu}}(E_{\nu}) = \frac{c n_0}{4 \pi} 
\int_0^{\rm z_{\rm max}} 
dz \psi(z) (1 + z) \left| \frac{dt}{dz} \right| 
\frac{d^2 N_{\nu}}{dt^s dE_{\nu}^s}(E_{\nu}^s, z), 
\label{eq:general_t}
\end{equation}
where $d^2 N_{\nu} / dt^s dE_{\nu}^s$ is the neutrino yield of a UHECR source located at the redshift of $z$ and $c$ is the speed of light. The superscript $s$ represents physical quantities measured at a redshift $z$. Throughout this paper, we call this frame the source frame. The comoving number density of UHECR sources, i.e., neutrino sources, is represented as $n_0 \psi(z)$ from the neutrino source number density in the local universe $n_0$ and its cosmological evolution factor $\psi(z)$. The evolution factor $\psi(z)$ is parametrized as $(1 + z)^m$ such that the parameter $m$ represents the ``scale'' of the cosmological evolution often used in the literature. Together with the maximum redshift of UHECR sources $z_{\rm max}$, this index describes a general characteristic of UHECR sources.

The neutrino yield is given by a convolution of the number of UHECRs, the number density of target photons, and the energy distribution of generated neutrinos per interaction. Cosmic rays are assumed to be distributed isotropically in the rest frame of the flow of plasma moving toward the observer with the Lorentz factor of $\Gamma$. Cosmic-ray distribution in the plasma rest frame (simply called the rest frame below) is Lorentz transformed to that in the source frame where cosmic rays are beamed. Now that cosmic rays are ultrarelativistic, neutrinos are well approximated to be produced in the same momentum directions as those of the cosmic rays, and then neutrinos are beamed. For convenience, we define the neutrino yield as an isotropic-equivalent form as 
\begin{equation}
\frac{d^2 N_{\nu}}{dt^s dE_{\nu}^s}(E_{\nu}^s) = 4 \pi \frac{d^3 N_{\nu}}{dt^s dE_{\nu}^s d\Omega_{\nu}^s}(E_{\nu}^s; \mu^s = 1), 
\end{equation}
where $d^3 N_{\nu} / dt^s dE_{\nu}^s d\Omega_{\nu}^s$ is an angular-dependent neutrino yield and $\mu^s$ is the cosine of the angle between the direction of the plasma flow and the momentum of neutrinos. The angular-dependent neutrino yield with $\mu^s = 1$ can be evaluated as a convolution of the number of cosmic rays moving along with the plasma motion, the number density of target photons, and the energy distribution of produced neutrinos per reaction. The energy distribution of cosmic rays toward the observer is represented as a power-law function if the spectrum of cosmic rays is described by a power-law function in the rest frame. Thus, we denote the spectrum of these cosmic rays in the source frame $d^2N_{\rm CR} / dt^s dE_{\rm CR}^s = \kappa_{\rm CR} (E_{\rm CR}^s / E_0^s)^{-\alpha}$. Here, $E_{\rm CR}^s$ is the energy of cosmic rays, and $E_0^s$ is the reference energy of cosmic rays in the source frame; it will be convenient to set $E_0^s = 10$ PeV when we consider PeV-energy neutrinos. The number of cosmic rays in the emission region can be estimated by using the escape time scale of cosmic rays from the region $t_{\rm esc}^s$ as $dN_{\rm CR} / dE_{\rm CR}^s = t_{\rm esc}^s d^2N_{\rm CR} / dt^s dE_{\rm CR}^s$. Note that this escape time scale is transformed into that in the rest frame $t'_{\rm esc}$ by $t'_{\rm esc} = \Gamma t_{\rm esc}^s$. Primed (') characters represent quantities measured in the rest frame below.

The energy distribution of neutrinos generated from an interaction is 
\begin{equation}
Y(E_{\nu}^s; E_{\rm CR}^s, s) = \int dE_{\pi} \frac{d \sigma_{\gamma p}}{dE_{\pi}^s}(E_{\pi}^s; E_{\rm CR}^s, s) \frac{d \rho_{\pi \rightarrow \nu}}{dE_{\nu}^s}(E_{\nu}^s; E_{\pi}^s), 
\label{eq:nudist}
\end{equation}
where $d \sigma_{\gamma p} / dE_{\pi}^s$ is the differential cross section of charged-pion production as a function of the energy of a generated charged pion $E_{\pi}^s$, $d \rho_{\pi \rightarrow \nu} / dE_{\nu}^s$ is the energy distribution of neutrinos produced by the decay of charged pions with the energy of $E_{\pi}^s$, and $s$ is a Mandelstam variable or the square of invariant mass of the cosmic-ray nucleon and a target photon in the photomeson collision with the cross section of $\sigma_{\gamma p}$. The approximated expressions of both the functions are given in Ref.~\cite{uhecr_constraint2012}.

The optical depth of $\gamma p$ interactions explicitly appears in the formula of the neutrino yield under our treatment. The energy distribution of neutrinos per interaction depends on only $E_{\rm CR}^s$ and $s$, not being dependent on the energy of a target photon $E_{\gamma}^s$ and its direction $\mu^s$ explicitly in Eq. (\ref{eq:nudist}). Furthermore, since the convolution of the neutrino energy distribution and the distribution of target photons is evaluated at the $\Delta$ resonance, it can be simplified to the multiplication of the neutrino energy distribution at $s = s_{\rm R} (\simeq 1.5~{\rm GeV}^2)$ and the inverse of the mean free path of cosmic rays with the energy of $E$, $\lambda_{\gamma p}(E)$. As the inverse of $\lambda_{\gamma p}(E)$ is the probability of interactions per unit path length, the optical depth of $\gamma p$ interactions, $\tau(E) = c t_{\rm esc} / \lambda_{\gamma p}(E)$, explicitly appears in the neutrino yield. Recalling that $\lambda_{\gamma
  p}(E)$ is transformed similarly to $t_{\rm esc}$, the optical depth can be evaluated in either the source frame or the rest frame. Thus, we evaluate the mean free path of $\gamma p$ interactions in the frame where target photons are distributed isotropically for simplicity.

Typically, there are two cases of target photon distribution. One is the case that photons are isotropically distributed in the rest frame of the plasma moving with the Lorentz factor $\Gamma$; i.e., the photons are distributed similarly to cosmic rays. This case corresponds to the situation that target photons are provided by electrons accelerated in the same place as cosmic rays, as mainly discussed for gamma-ray bursts and relatively weak blazars, that is, BL Lac objects. The other case is that photons are isotropically distributed in the source frame. This treatment is appropriate if target photons are mainly provided outside the relativistic plasma flow. This situation is often realized in quasars. Below we derive the analytical expression of diffuse neutrino intensity in the first case, but we can also obtain that in the second case by replacing several quantities, as will be shown afterwards.

The mean free path of $\gamma p$ interactions is given by 
\begin{equation}
\frac{1}{\lambda_{\gamma p}(E'_{\rm CR})} = \int ds \frac{dn_{\gamma}}{ds} \sigma_{\gamma p}(s). 
\end{equation}
We approximate the energy distribution of isotropically distributed target photons by a power-law function, $dn_{\gamma} / dE'_{\gamma} = \kappa' {E'_{\gamma}}^{-\gamma}$ in the range of ${E'_{\gamma}}^{\rm min} \leq E'_{\gamma} \leq {E'_{\gamma}}^{\rm max}$, for a model-independent approach. The energy range of protons contributing to photomeson production is typically $(s_{\rm R} - m_p^2) / 4 {E'_{\gamma}}^{\rm max} \lesssim E'_{\rm CR} \lesssim (s_{\rm R} - m_p^2) / 4 {E'_{\gamma}}^{\rm min}$, where $m_p$ is the proton mass. Since the amount of cosmic rays with the energy of $E'_{\rm CR} > (s_{\rm R} - m_p^2) / 4 {E'_{\gamma}}^{\rm min}$ is small due to their steep spectrum, this energy range is negligible in the integration over $E'_{\rm CR}$. On this assumption, the spectral number density of photons is written as 
\begin{equation}
\frac{dn_{\gamma}}{ds} = \frac{\kappa'}{4 E'_{\rm CR}} \left( \frac{s - m_p^2}{2 E'_{\rm CR}} \right)^{-\gamma} \frac{2^{\gamma + 1}}{\gamma + 1}, 
\end{equation}
as long as $E_{\rm CR}^s > \Gamma (s_{\rm R} - m_p^2) / 4 {E'_{\gamma}}^{\rm max}$. Otherwise, the number density is zero. Here, we use $E_{\rm CR}^s = \Gamma E'_{\rm CR}$, which is justified by the fact that we focus on only cosmic rays directed to the observer. Thus, the mean free path is represented as 
\begin{equation}
\frac{1}{\lambda_{\gamma p}(E'_{\rm CR})} = \frac{\kappa'}{4 E'_{\rm CR}} 
\frac{2^{\gamma + 1}}{\gamma + 1} \int ds \sigma_{\gamma p}(s) 
\left( \frac{s - m_p^2}{2 E'_{\rm CR}} \right)^{-\gamma}. 
\end{equation}
This indicates that the energy dependence of the optical depth is 
\begin{equation}
\tau_{\gamma p}(E_{\rm CR}^s) = \tau_0 \left( \frac{E_{\rm CR}^s}{E_0^s} \right)^{\gamma - 1}, 
\label{eq:opt_depth}
\end{equation}
with $\tau_0$, the optical depth at $E_{\rm CR}^s = E_0^s$, assuming that the escape time scale is independent of the energy of cosmic rays.

In fact, the escape mechanism of cosmic rays from a neutrino production region is not clear yet. While the advective escape time scale is independent of cosmic-ray energy, the diffusive escape time scale depends on cosmic-ray energy. Even if the escape time scale depends on the cosmic-ray energy with a power-law form, Eq. (\ref{eq:opt_depth}) can be applied after reinterpreting $\gamma$ as the sum of the target-photon spectral index and the index of the escape time scale.

Using the expression of the optical depth, the isotropic equivalent neutrino yield of a UHECR source is 
\begin{equation}
\frac{d^2 N_{\nu}}{dt^s dE_{\nu}^s}(E_{\nu}^s) \simeq 
\int dE_{\rm CR}^s \kappa_{\rm CR} \left( \frac{E_{\rm CR}^s}{E_0^s} \right)^{-\alpha} 
Y(E_{\nu}^s; E_{\rm CR}^s, s_{\rm R}) \tau(E_{\rm CR}^s). 
\end{equation}

The intensity of diffuse neutrinos is obtained by substituting the neutrino yield into Eq. (\ref{eq:general_t}). If we focus on leading terms for simplicity after integrating the formula by $z$ with the approximation in Appendix \ref{sec:app1}, the neutrino intensity is 
\begin{widetext}
\begin{equation}
\frac{d J_{\nu}}{dE_{\nu}}(E_{\nu}) \simeq 
\frac{n_0 \kappa_{\rm CR} \tau_0}{(\alpha + 1 - \gamma)^2} \frac{c}{H_0} 
\frac{s_{\rm R}}{\sqrt{(s_{\rm R} + m_{\pi}^2 - m_p^2)^2 - 4 s_{\rm R} m_{\pi}^2}} 
\frac{3}{1 - r_{\pi}} {E_0^s}^{\alpha + 1 - \gamma} \zeta, 
\label{eq:onsource_final}
\end{equation}
\end{widetext}
where $r_{\pi} = m_{\mu}^2 / m_{\pi}^2 \simeq 0.57$ is the muon-to-pion mass-squared ratio, and the factor of 3 corresponds to the number of neutrinos produced from the $\pi$ meson and $\mu$ lepton decay chain. The factor $\zeta$ in Eq. (\ref{eq:onsource_final}) is the term that accounts for the redshift dependence and is given by, 
\begin{widetext}
\begin{eqnarray}
\zeta &=& I(z_{\rm down}, z_{\rm up}) \left( \frac{E_{\nu}}{x_{\rm R}^+ (1 - r_{\pi})} \right)^{-(\alpha + 1 - \gamma)} 
- I(\tilde{z}_{\rm down}, z_{\rm down}) \left( \frac{E_{\nu}}{x_{\rm R}^- (1 - r_{\pi})} \right)^{-(\alpha + 1 - \gamma)} \nonumber \\
&& ~~~~~~~~~~~~~~~~~~~~~~~~~~~~~~~~~~~~~~~~~~~~~~~~~~~~~~~~~~ 
+ \frac{2}{2 m - 1} \Omega_{\rm M}^{- \frac{m + 1}{3}} 
\left( \Gamma \frac{s_{\rm R} - m_p^2}{4 {E'_{\gamma}}^{\rm max}} \right)^{- (\alpha + 1 - \gamma)} \Upsilon \label{eq:zeta} \\
I(z_1, z_2) &=& \frac{2}{2 (m - \alpha + \gamma) - 3} \Omega_{\rm M}^{- \frac{m - \alpha + \gamma}{3}} 
\left[ \left\{ \Omega_{\rm M} (1 + z_2)^3 + \Omega_{\Lambda} \right\}^{\frac{m - \alpha + \gamma}{3} - \frac{1}{2}} - \left\{ \Omega_{\rm M} (1 + z_1)^3 + \Omega_{\Lambda} \right\}^{\frac{m - \alpha + \gamma}{3} - \frac{1}{2}} \right], \\
\Upsilon &=& 
\left\{ \Omega_{\rm M} (1 + z_{\rm down})^3 + \Omega_{\Lambda} \right\}^{\frac{m}{3} - \frac{1}{6}} - \left\{ \Omega_{\rm M} (1 + \tilde{z}_{\rm down})^3 + \Omega_{\Lambda} \right\}^{\frac{m}{3} - \frac{1}{6}} - (\alpha + 1 - \gamma) \log \left( \frac{x_{\rm R}^+}{x_{\rm R}^-} \right) \nonumber \\
&& ~ + (\alpha + 1 - \gamma) 
\left\{ \Omega_{\rm M} (1 + z_{\rm down})^3 + \Omega_{\Lambda} \right\}^{\frac{m}{3} - \frac{1}{6}} \left[ \ln \left( \frac{s_{\rm R} - m_p^2}{4 E_{\nu} (1 + z_{\rm down})} \frac{\Gamma}{{E'_{\gamma}}^{\rm max}} x_{\rm R}^+ (1 - r_{\pi}) \right) + \frac{2}{2 m - 1} \right] \nonumber \\
&& ~ - (\alpha + 1 - \gamma) 
\left\{ \Omega_{\rm M} (1 + \tilde{z}_{\rm down})^3 + \Omega_{\Lambda} \right\}^{\frac{m}{3} - \frac{1}{6}} \left[ \ln \left( \frac{s_{\rm R} - m_p^2}{4 E_{\nu} (1 + \tilde{z}_{\rm down})} \frac{\Gamma}{{E'_{\gamma}}^{\rm max}} x_{\rm R}^+ (1 - r_{\pi}) \right) + \frac{2}{2 m - 1} \right], 
\label{eq:upsilon}
\end{eqnarray}
\end{widetext}
where $x_{\rm R}^{\pm}$ are the maximal and minimal bounds of the relative energy of the emitted pion normalized by the parent cosmic-ray energy. These are given by a kinematical relation, 
\begin{equation}
x_{\rm R}^{\pm} = \frac{(s_{\rm R} + m_{\pi}^2 - m_p^2) \pm \sqrt{(s_{\rm R} + m_{\pi}^2 - m_p^2)^2 - 4 s_{\rm R} m_{\pi}^2}}{2 s_{\rm R}}. 
\end{equation}
It is worth noting that the main energy range of neutrinos produced by an UHECR nucleon with the energy of $E_{\rm CR}$ is given by
\begin{equation}
E_{\rm CR} x_{\rm R}^-(1-r_\pi)\lesssim E_\nu \lesssim E_{\rm CR} x_{\rm R}^+(1-r_\pi).
\label{eq:neut_cr_energy_range}
\end{equation}

The characteristic redshifts $z_{\rm up}$, $z_{\rm down}$, and $\tilde{z}_{\rm down}$ that appear in Eqs.~(\ref{eq:zeta}) and (\ref{eq:upsilon}) depend on neutrino energies $E_{\nu}$ due to kinematics of $\pi$ decay and the redshift energy loss. The expression of $z_{\rm up}$ is given by
\begin{widetext}
\begin{equation}
1 + z_{\rm up} = \left\{
\begin{array}{lc}
1 + z_{\rm max} & \left( E_{\nu} < \frac{s_{\rm R} - m_p^2}{4 (1 + z_{\rm max})} \frac{\Gamma}{{E'_{\gamma}}^{\rm min}} x_{\rm R}^- (1 - r_{\pi}) \right) \\
\frac{s_{\rm R} - m_p^2}{4} \frac{\Gamma}{{E'_{\gamma}}^{\rm min}} \frac{x_{\rm R}^- (1 - r_{\pi})}{E_{\nu}} & \left( \frac{s_{\rm R} - m_p^2}{4 (1 + z_{\rm max})} \frac{\Gamma}{{E'_{\gamma}}^{\rm min}} x_{\rm R}^- (1 - r_{\pi}) \leq E_{\nu} \leq \frac{s_{\rm R} - m_p^2}{4} \frac{\Gamma}{{E'_{\gamma}}^{\rm min}} x_{\rm R}^- (1 - r_{\pi}) \right) \\
1 & \left( \frac{s_{\rm R} - m_p^2}{4} \frac{\Gamma}{{E'_{\gamma}}^{\rm min}} x_{\rm R}^- (1 - r_{\pi}) < E_{\nu} \right)
\end{array}
\right.
\label{eq:redshift_up}
\end{equation}
\end{widetext}
The other two characteristic redshifts can be obtained as follows: $z_{\rm down}$ is also given by Eq.~(\ref{eq:redshift_up}) replacing ($x^-_{\rm R}$, ${E'_\gamma}^{\rm min}$) by ($x^+_R$, ${E'_\gamma}^{\rm max}$), and $\tilde{z}_{\rm down}$ is given by Eq.~(\ref{eq:redshift_up}) replacing ${E'_\gamma}^{\rm min}$ by ${E'_\gamma}^{\rm max}$.

The ${E'_\gamma}^{\rm max/min}$ and $x_{\rm R}^{\pm}$ dependence of $\zeta$ reflects the association between the parent cosmic-ray energy $E_{\rm CR}$ and the target photon energy $E_\gamma$in the photomeson collision. When $E_{\nu} \gtrsim (s_{\rm R} - m_p^2) \Gamma x_{\rm R}^- (1 - r_{\pi}) / 4 {E'_{\gamma}}^{\rm min}$, $\zeta \sim 0$ and the neutrino flux thus becomes null. This is a consequence of the fact that there are no target photons that can interact with cosmic rays through the $\Delta$ resonance. The energies of target photons are within the photomeson collision range in all sources distributed from $z=0$ to $z=z_{\rm max}$ when the neutrino energy satisfies the following condition:
\begin{equation}
\frac{s_{\rm R} - m_p^2}{4 {E'_{\gamma}}^{\rm max}} \Gamma x_{\rm R}^+ (1 - r_{\pi}) \lesssim E_{\nu} \lesssim \frac{s_{\rm R} - m_p^2}{4 {E'_{\gamma}}^{\rm min} (1 + z_{\rm max})} \Gamma x_{\rm R}^- (1 - r_{\pi}). 
\label{eq:neut_energy_range}
\end{equation}
The neutrino flux given by Eq.~(\ref{eq:onsource_final}) follows the simple power law of $\sim E_\nu^{-\alpha+\gamma-1}$ in this range. If we further assume $\gamma=1$, {\it i.e.}, the optical depth does not depend on cosmic-ray energies [see Eq.~(\ref{eq:opt_depth})], the on-source neutrino flux can be represented by the simpler form
\begin{widetext}
\begin{eqnarray}
\frac{d J_{\nu}}{dE_{\nu}}(E_{\nu}) &\simeq& 
\frac{2 n_0 \kappa_{\rm CR}}{\alpha^2} \frac{c}{H_0} 
\frac{s_{\rm R}}{\sqrt{(s_{\rm R} + m_{\pi}^2 - m_p^2)^2 - 4 s_{\rm R} m_{\pi}^2}} 
\frac{3}{1 - r_{\pi}} 
\frac{{E_0^s}^{\alpha} (1 - e^{-\tau_0})}{2 (m - \alpha) - 1} \Omega_{\rm M}^{- \frac{m - \alpha + 1}{3}} \nonumber \\
&& ~~~~~~~~~~~~~~~~~~~~~~~~~~~~~~~ \times 
\left[ \left\{ \Omega_{\rm M} (1 + z_{\rm max}) + \Omega_{\Lambda} \right\}^{\frac{m - \alpha}{3} - \frac{1}{6}} - 1 \right] 
\left( \frac{E_{\nu}}{x_{\rm R}^+ (1 - r_{\pi})} \right)^{- (\alpha + 1 - \gamma)}. 
\label{eq:onsource_simple}
\end{eqnarray}
\end{widetext}
Here, we replace $\tau_0$ by $1 - e^{-\tau_0}$ to treat the case with $\tau_0 \sim 1$. This equation corresponds to an advanced formulation of the pioneering work known as the Waxman-Bahcall limit~\cite{waxman98} based on a simplified energetics argument to estimate an upper bound of the on-source neutrino flux on the simple assumption of $\gamma = 1$. It is worth noticing that the expressions of the diffuse neutrino intensity, i.e., Eqs. (\ref{eq:onsource_final}) and (\ref{eq:onsource_simple}), are not strongly dependent on the Lorentz factor of the plasma $\Gamma$ explicitly because the effects of $\Gamma$ are included in the definitions of some parameters, such as $\kappa_{\rm CR}$.

\begin{figure}[tb]
  \includegraphics[width=0.4\textwidth]{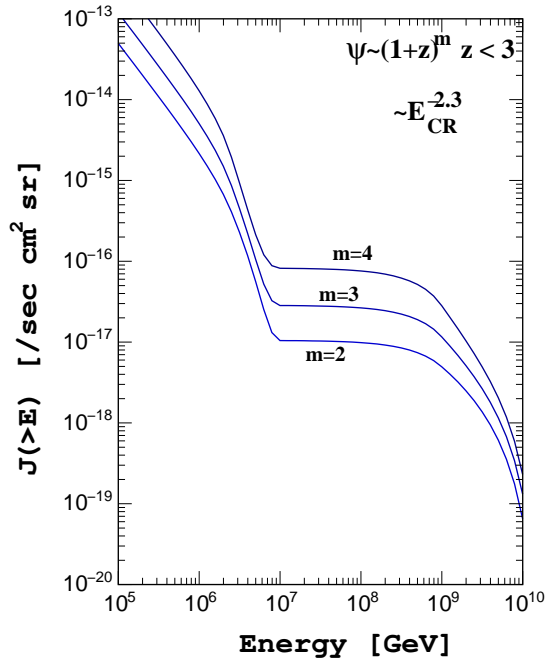}
  \caption{Integral neutrino fluxes, $J(>E)$ [cm$^{-2}\ \sec^{-1}$ sr$^{-1}$], as a function of neutrino energy. Bold lines represent the present analytical estimates with the various evolution factor $m$. The fluxes above $10^7$~GeV are dominantly generated by the GZK cosmogenic neutrinos calculated by the analytical formula in Ref.~\cite{uhecr_constraint2012}.
\label{fig:integral_fluxes}} 
\end{figure}

So far, we have considered the case that target photons are isotropically distributed in the rest frame of the plasma moving with the Lorentz factor $\Gamma$. In the case that photons are isotropically distributed in the source frame, the diffuse neutrino intensity can be obtained by removing $\Gamma$ in Eqs. (\ref{eq:zeta}), (\ref{eq:upsilon}), and (\ref{eq:redshift_up}) and replacing ${E'_{\gamma}}^{\rm min / max}$ into $E_{\gamma}^{\rm min / max}$, which are the minimum and maximum energy of target photons in the source frame.

Figure~\ref{fig:integral_fluxes} displays the neutrino fluxes calculated by the present analytical equation for the on-source neutrinos and the formula in Ref.~\cite{uhecr_constraint2012} for the GZK cosmogenic neutrinos. Here, $\alpha=2.3$, $\gamma=1$, $\tau_0=0.1$, $\Gamma=1$,$z_{\rm max}=3$, and ${E'_\gamma}^{\rm min}=0.3$~eV are assumed. The absolute intensity determined by $n_0$ and $\kappa_{\rm CR}$ was calculated from the observed UHECR flux described in Ref.~\cite{uhecr_constraint2012} for illustrative purposes. The on-source fluxes have a smooth cutoff at around 3 PeV in this particular set of realizations because ${E'_\gamma}^{\rm min} = 0.3$~eV implies that a major population of neutrinos is in energy range of $E_\nu \lesssim (s_{\rm R} - m_p^2) \Gamma x_R^{-} (1 - r_\pi) / (4 (1+z_{\rm max}){E'_\gamma}^{\rm min})\sim 3$~PeV as described by Eq.~(\ref{eq:neut_energy_range}). In the present formulation, the parameters $E_\gamma^{\rm min}$ and $\Gamma$ control the maximal
  energy of neutrinos. Other factors such as synchrotron cooling of secondary muons could set the maximal energy bound as well, although they are not taken into account in our calculations. However, the mechanism setting the maximal energy of neutrinos is irrelevant in constraining UHECR source characteristics if the only arguments are based on the total neutrino intensity and the event rate seen by IceCube. It is insensitive to the detailed spectral shape of neutrinos such as the cut-off structure.

Hereafter we fix $\Gamma = 1$ and $\gamma = 1$ for simplicity and clearness. The overall conclusion remains valid in changing these parameters.

\section{\label{sec:result}Results}
\subsection{\label{subsec:optdepth}
The constraints on the optical depth and extragalactic UHECR nucleon flux}

The final formula of a neutrino flux given by Eq.~(\ref{eq:onsource_final}) is linked with the intensity of extragalactic cosmic rays through the cosmic-ray yield $\kappa_{\rm CR}$. The UHECR intensity is calculated by integrating a UHECR spectrum over all the sources in the redshift space as 
\begin{widetext}
\begin{equation}
\frac{d J_{\rm CR}}{dE_{\rm CR}}(E_{\rm CR}) = 
n_0 c \kappa_{\rm CR} \int_0^{z_{\rm max}} 
(1 + z)^{1 - \alpha} \psi(z) \left| \frac{dt}{dz} \right| 
e^{- \tau_0} 
\left( \frac{E_{\rm CR}}{E_0^s} \right)^{-\alpha}, 
\label{eq:UHECR_flux}
\end{equation}
\end{widetext}
with neglecting intergalactic magnetic fields and the energy-loss in the CMB field during UHECR propagation. The convolution of Eq.~(\ref{eq:onsource_final}) with the effective area of IceCube~\cite{icecubePeV2013,icecubeHESE2013,icecubeEHE2013} predicts a neutrino event rate. Compared with the allowed range of the number of signal neutrino events in the IceCube observations estimated by the Feldman-Cousins (FC) upper and lower bounds~\cite{feldman98}, the predicted event rate places constraints on extragalactic UHECR intensity and the optical depth $\tau_0$.

\begin{figure}[tb]
  \includegraphics[width=0.4\textwidth]{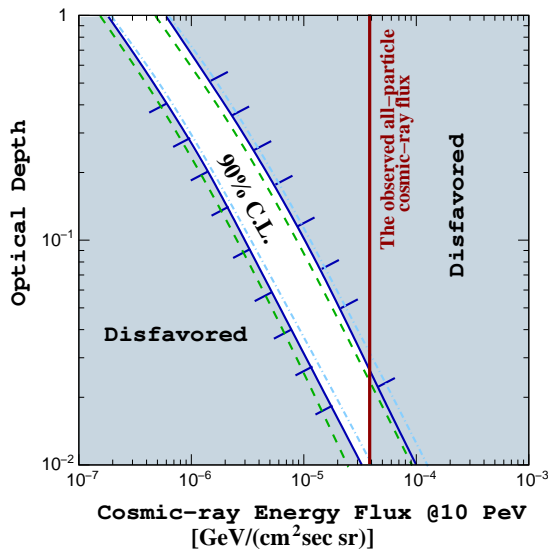}
  \caption{Constraints on the optical depth of UHECR sources for PeV-energy neutrino production and the energy flux of extragalactic UHECRs $E_{\rm CR}^2dJ_{\rm CR}/dE_{\rm CR}$ at energy of 10 PeV. The region sandwiched by the two blue solid curves ($\alpha=2.5$), green dashed curves ($\alpha=2.7$) and light blue dot-dashed curves ($\alpha=2.3$) is allowed by the present IceCube observations~\cite{icecubePeV2013,icecubeHESE2013}, respectively. The unshaded region highlights the allowed region for $\alpha=2.5$, taking into account the observed intensity of UHECRs measured by the IceTop experiment~\cite{icetop2013}.
\label{fig:constraints_10PeV}} 
\end{figure}

\begin{figure*}[bt]
  \includegraphics[width=0.4\textwidth]{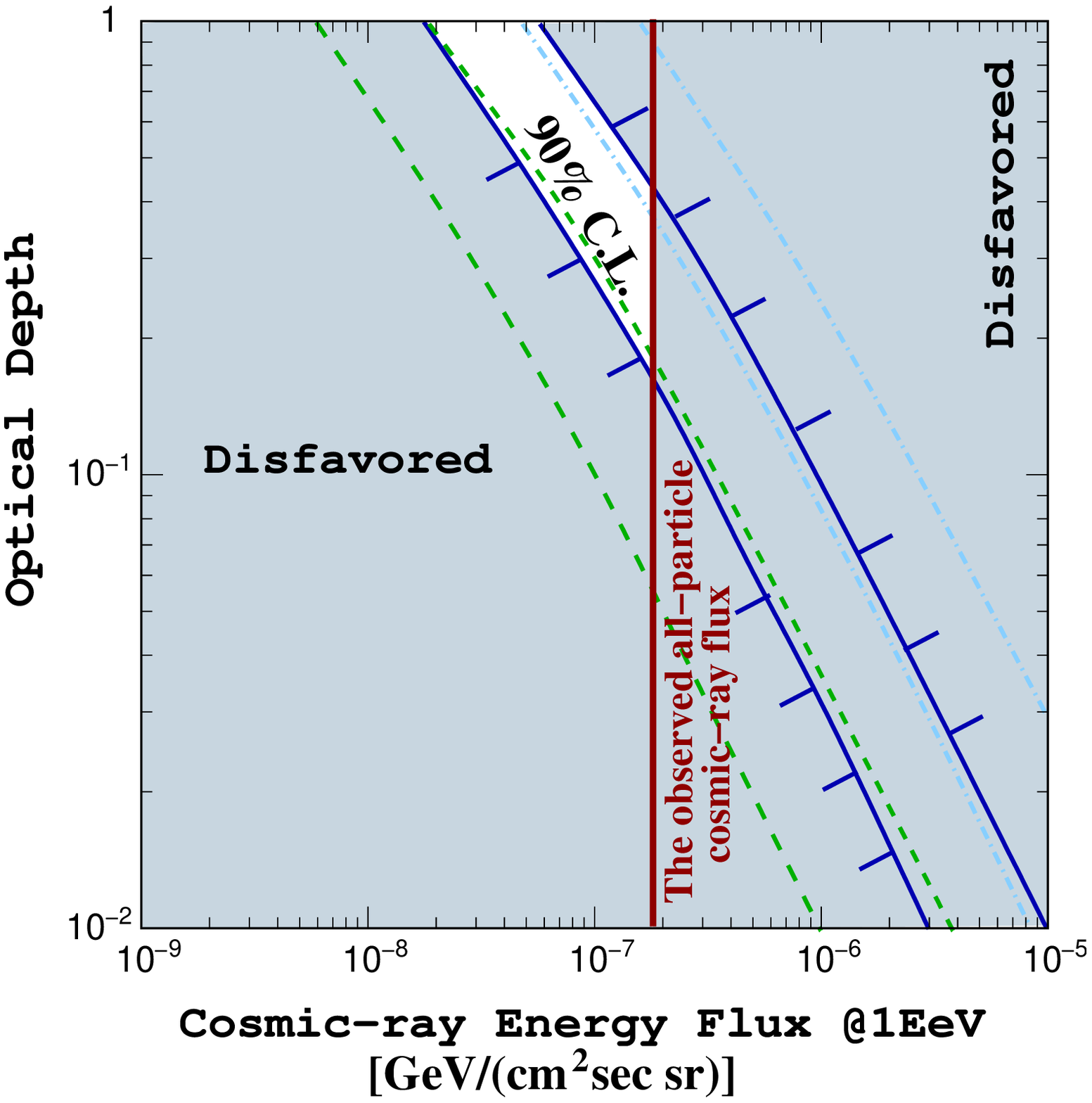}
  \includegraphics[width=0.4\textwidth]{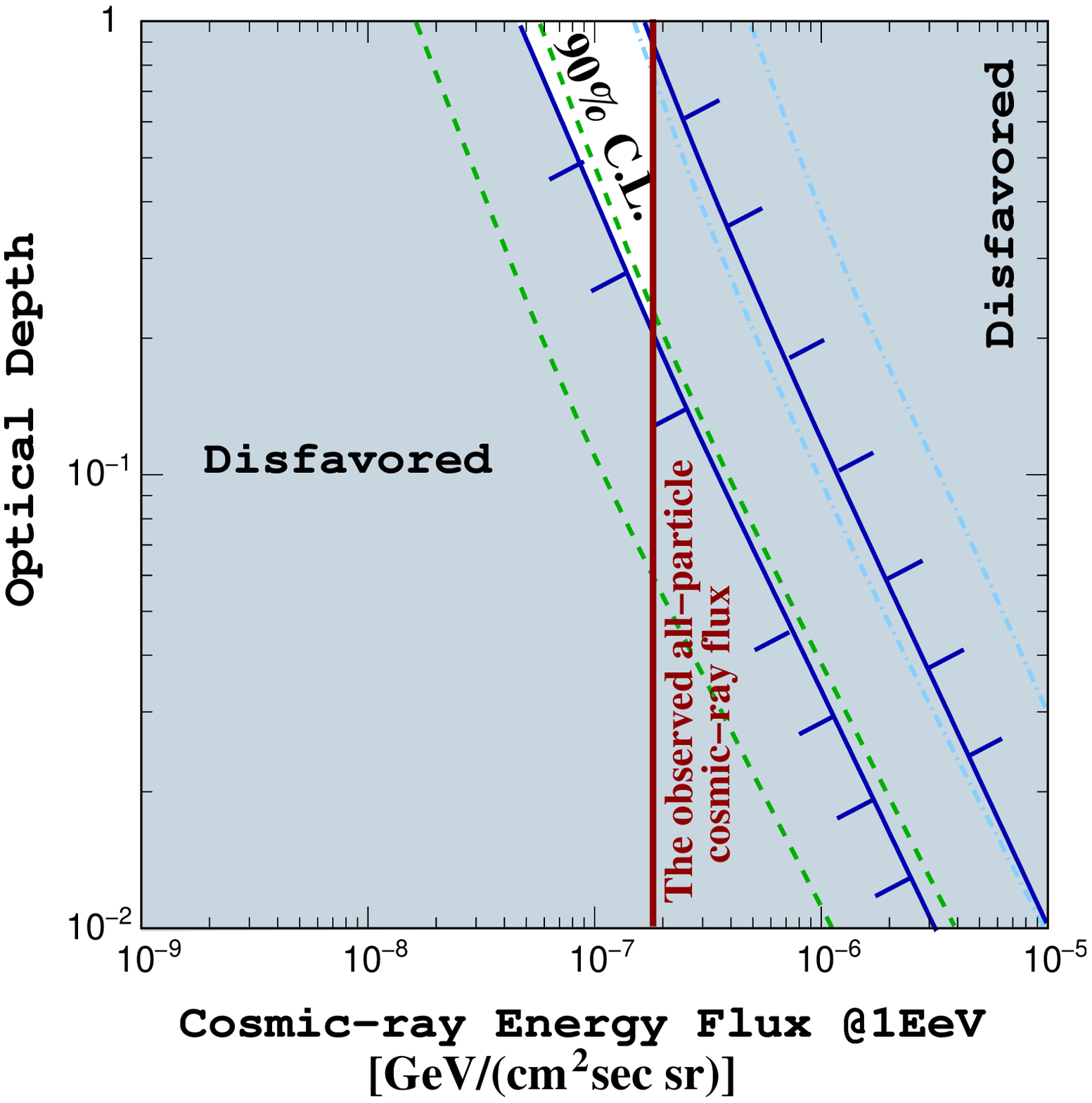}
  \caption{Same as Fig.~\ref{fig:constraints_10PeV}, but constraints against the cosmic ray flux at energy of 1 EeV. It displays the constraints when UHECR spectrum from the PeV neutrino sources extends to higher energies. Two cases are shown when the beam dumping factor $e^{-\tau_0}$ is included (left panel) or not included (right panel) in the calculation of the UHECR flux at 1 EeV.
\label{fig:constraints_1EeV}} 
\end{figure*}

First, we constrain these parameters at 10 PeV, {\it i.e.}, $10^7$~GeV, the representative cosmic-ray energy to produce PeV-energy neutrinos by $\gamma p$ interactions. The range of the number of astrophysical neutrino events $N_\nu$ at the FC 90 \% confidence level is $8.0 \leq N_\nu \leq  26.0$ for the IceCube observation of the 28 event detection with the expected background of 12.0 in the energy region of $E_\nu \gtrsim 3\times 10^4$~GeV~\cite{icecubeHESE2013}. The intensity of neutrinos per flavor is treated as one-third of the all-flavor neutrino intensity estimated by the analytical function, assuming the full mixing in the standard neutrino oscillation scenario. Figure~\ref{fig:constraints_10PeV} displays the resultant constraints on the optical depth and the extragalactic cosmic-ray proton intensity for several values of $\alpha$, all of which are consistent with the IceCube observation at the present statistics~\cite{icecubeHESE2013}.The parameters concerning the UHECR SDF, $m$ and $z_{\rm max}$, are assumed to follow the star formation rate~\cite{hopkins2006} here and are parametrized as $\psi(z_{\rm s}) \propto (1 + z_{\rm s})^m$ where~\cite{kotera2010,uhecr_constraint2012}
\begin{eqnarray}
\psi(z_{\rm s}) \propto \left\{ 
\begin{array}{ll}
(1 + z_{\rm s})^{3.4} & ( 0 \leq z_{\rm s} \leq 1 ) \\
{\rm const} & (1 \leq z_{\rm s} \leq 4)
\end{array}
\right. .
\label{eq:sdf}
\end{eqnarray}
We should remark that these constraints do not strongly depend on an assumed SDF because it appears in both the neutrino and UHECR fluxes and its effect is almost canceled. The allowed region in the parameter space is rather small and is almost independent of the spectral index of cosmic rays $\alpha$. The optical depth of $\gamma p$ interactions needed to reproduce the PeV-energy neutrino detection is $\tau_0 \gtrsim 10^{-2}$. The optical depth should be less than unity for UHECRs producing the PeV-energy neutrinos and also contributing to the observed total cosmic-ray flux. Under this requirement, extragalactic cosmic rays must occupy at least a few percent of the observed total cosmic-ray flux.

An interesting possibility is that the spectrum of UHECRs producing the PeV-energy neutrinos is further extended to higher energy. Assuming that the spectrum of such UHECRs is extended above $1$ EeV with a simple power-law form, we derive constraints on the $\gamma p$ optical depth by using the total cosmic-ray flux at $1$ EeV, {\it i.e.}, $10^9$~GeV, shown in Fig.~\ref{fig:constraints_1EeV}. The allowed regions in the parameter space become much smaller than in the previous case because the spectrum of the total cosmic rays is steeper than that of the extragalactic UHECR protons with the power law index of $\alpha$ whose range is bounded by the IceCube observation. Note that $\tau_0$ constrained here is the $\gamma p$ optical depth of $E_0^s = 10$ PeV protons.

The dumping factor $e^{-\tau_0}$ [see Eq.~(\ref{eq:UHECR_flux})] is taken into account to calculate the UHECR flux at 1 EeV in the left panel of Fig.~\ref{fig:constraints_1EeV}. In other words, we consider the situation that $\gamma p$ interactions also occur for UHECRs with the energies of EeV, that is, EeV-energy neutrinos are also produced. In this case some mechanisms to suppress the resultant neutrino flux at EeV energies are likely to be required to be reconciled with the fact that IceCube has not seen EeV-energy neutrinos in spite of its larger effective area than that at PeV energies~\cite{icecubeEHE2013}, e.g., the synchrotron cooling of $\pi / \mu$ in a strongly magnetized region before neutrino emission. The other extreme case is that the $\gamma p$ collision does not occur in the EeV range, shown in the right panel of Fig.~\ref{fig:constraints_1EeV} where the dumping factor is dropped from the UHECR flux calculation. In either case, the constraints suggest that the optical depth of protons for PeV-energy neutrinos is rather high, $\tau_0 \gtrsim 0.2$, and also that a major fraction of UHECRs in the EeV region is extragalactic protons. It is supportive to the ``dip'' transition model~\cite{berezinsky2006} where the {\it ankle} structure of the cosmic-ray spectrum, which appears at 3 to 10\,EeV, is caused by the energy loss of UHECR protons by the Bethe-Heitler pair production with CMB photons. The dip transition model also predicts high cosmogenic neutrino flux up to $10^8$ GeV~\cite{takami09,kotera2010}.

\subsection{\label{subsec:evolution}
The constraints on the UHECR source evolution}

\begin{figure}[bt]
\includegraphics[width=0.4\textwidth]{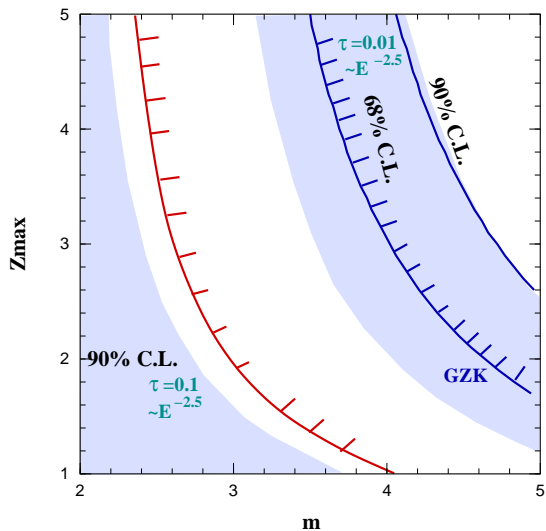}
\caption{Constraints on the UHECR source evolution index $m$ and the maximum redshift of the sources $z_{\rm max}$ with the IceCube observations. The area above the blue curves is excluded with the null detection of neutrinos with energies above 100 PeV at the quoted confidence levels~\cite{icecubeEHE2013} by requiring the GZK cosmogenic neutrino flux to be consistent with the null detection. The shaded region displays the allowed parameter space by the detection of PeV-energy neutrinos assuming that the sources of the highest energy cosmic rays also produce the PeV-energy neutrinos. The two representative cases ($\tau_0=0.1$ and $0.01$) are shown for $\alpha=2.5$. The area below the red curve is consistent with the total cosmic-ray flux at 1 EeV.}
\label{fig:gzk_constraint}
\end{figure}

The expected event rates with the IceCube neutrino observatory provides constraints on the cosmological evolution of UHECR sources. Above 100 PeV, GZK cosmogenic neutrino production is considered as a ``guaranteed'' mechanism to yield neutrinos. The null detection of neutrinos above 100 PeV by IceCube has put solid bounds on the source evolution parameters $m$ and $z_{\rm max}$~\cite{icecubeEHE2013} with the analytical formulation of the cosmogenic neutrino flux~\cite{uhecr_constraint2012}. The blue curves in Fig.~\ref{fig:gzk_constraint} shows the obtained bounds assuming that all the UHECRs in the highest energy region are protons.

Here we further constrain the evolution parameter space by using the derived analytical formulae assuming that the sources of the highest energy cosmic rays also produce the on-source PeV-energy neutrinos that IceCube has detected. The normalization factor of UHECR flux, $\kappa_{\rm CR}$, is determined by the observed integral flux at $E_{\rm CR}$ of 100 EeV as in Ref.~\cite{uhecr_constraint2012} in this case. With this condition, the present analytical formula of neutrino flux gives the event rate as a function of $m$ and $z_{\rm max}$ under a fixed $\tau_0$. The neutrino flux is compared to the observed event rate with the FC statistical confidence level (see the previous section). The shaded area in Fig.~\ref{fig:gzk_constraint} is the allowed region in the parameter space obtained from this comparison under two representative $\gamma p$ optical depths $\tau_0 = 0.1$ and $0.01$. The red curve is the boundary for the calculated cosmic-ray flux not to exceed the total cosmic
 -ray flux at 1 EeV.

Fig.~\ref{fig:gzk_constraint} indicates that the cases of $\tau_0 = 10^{-2}$ are already ruled out by the constraints from the intensity of cosmic-ray flux at 1 EeV because strong cosmological evolution, i.e., large $m$ with high $z_{\rm max}$, is required. About a half of this parameter space is also disfavored by the null detection of cosmogenic neutrinos by IceCube. An optically thicker source class with $\tau_0 = 0.1$ is consistent with these two constraints. Thus, if the sources of PeV-energy neutrinos are directly connected to the sources of the highest energy cosmic rays, they have relatively large $\gamma p$ optical depth to produce PeV-energy neutrinos and relatively weak cosmological evolution (see Sec. \ref{sec:case_studies} for discussions on specific source classes). Note that the results depend on the spectral index $\alpha$.

\section{\label{sec:discussion} Discussion}

\subsection{\label{sec:case_studies} Case study for specific astronomical objects}

We have obtained the general constraints on the optical depth of $\gamma p$ interactions to produce PeV-energy neutrinos. The connection between the IceCube neutrino flux and cosmic-ray flux at 10 PeV indicates the $\gamma p$ optical depth of $\tau_0 \gtrsim 10^{-2}$ and the fraction of extragalactic cosmic ray protons in the observed total cosmic ray intensity above a few percent (Fig. \ref{fig:constraints_10PeV}). If the spectrum of cosmic rays producing the PeV-energy neutrinos extends up to 1 EeV, the constraints on the optical depth become more stringent, $\tau_0 \gtrsim 0.2$, and require that more than 10\% of extragalactic cosmic-ray protons should be responsible for the total cosmic-ray intensity at 1 EeV (Fig. \ref{fig:constraints_1EeV}). Furthermore, if the cosmic-ray spectrum extends to the highest energies and explains the observed flux of UHECRs with the energies of 100 EeV, i.e., the spectrum is normalized at this energy, the cosmological evolution of UHECRs is constrained on the $m$ - $z_{\rm max}$ plane (Fig. \ref{fig:gzk_constraint}). In this section we survey PeV-energy neutrino emission models for specific astronomical objects, i.e., active galactic nuclei (AGN) and gamma-ray bursts (GRBs), and check compatibility with these constraints.

\subsubsection{Active galactic nuclei}

AGNs are one of the promising sites to accelerate UHECRs \cite{UHECR_AGN} and therefore are expected to produce high-energy neutrinos. Several classical predictions of neutrino flux have been already ruled out by AMANDA-II and IceCube~\cite{nu_ruledout}. Surviving models are mainly related to blazars, which are radio-loud AGN with relativistic jets directed toward us. Hence, here we focus on blazars as sources of high-energy neutrinos.

Blazars are classified into two categories by their activity: BL Lac objects and flat spectrum radio quasars (FSRQs). In both cases, particles are believed to be accelerated in inner jets, subparsec-to-parsec scale from the central supermassive black hole. FSRQs have accretion disks with activity higher than BL Lac objects, and therefore the particle acceleration region of FSRQs is filled by photons from the accretion disks, broad line regions, and dusty tori. These are the main target photons for photomeson production to produce neutrinos and also for inverse Compton scattering of electrons to produce gamma rays with leptonic origin. On the other hand, since BL Lac objects have only weak external photon fields, target photons are only synchrotron radiation of electrons accelerated in the same region as UHECRs. The origin of gamma rays has been studied to identify the acceleration of UHECRs, but the hadronic origin of gamma rays has never been identified clearly; the spectrum of blazars can be reproduced by both leptonic \cite{Maraschi1992ApJ397L5,Dermer1993ApJ416p458} and hadronic models \cite{Mannheim1993A&A269p67,Mucke2003Aph18p593} at present. High-energy neutrinos are an alternative, powerful way to confirm the acceleration of UHECRs in blazars.

BL Lac objects have very small $\gamma p$ optical depth $\tau_0$ because of their weak synchrotron power. In the cases that gamma rays are of leptonic origin, a recent observational result with good spectral coverage for a quiescent state of Mrk 421 \cite{Abdo2011ApJ736p131} allows us to estimate the optical depth $\tau_0$ of $\lesssim 10^{-7}$ if protons are accelerated in the same region as electrons. The optical depth is $\lesssim 10^{-6}$ even in flaring episodes. Hadronic models for gamma rays require much stronger magnetic fields in the acceleration/emission region, $10$ -- $100$ G, than those in the leptonic model, typically $\sim 0.1$ G, to accelerate protons up to UHE, in order to open the $\gamma p$ channel and also to lead protons to emit synchrotron radiation in gamma-ray bands. However, because of small electron-synchrotron photon density, proton synchrotron radiation dominantly contributes to gamma rays, and therefore the optical depth $\tau_0$ is still $\sim 10^{-7}$ for Mrk 421. Even considering a relatively bright BL Lac object, PKS 0716+714, i.e., a member of low-frequency-peaked BL Lac objects (LBLs), the optical depth $\tau_{p\gamma}$ reaches $\sim 3 \times 10^{-4}$ \cite{Mucke2003Aph18p593}. Hence, the optical depth is much smaller than the requirement displayed in Fig.~\ref{fig:constraints_10PeV}, indicating that BL Lac objects do not dominantly contribute to the PeV-energy neutrino flux.

FSRQs can realize a much larger $\gamma p$ optical depth because of strong external photon fields and their own high luminosity. Especially photons in broad line regions contribute to the production of PeV neutrinos as long as a proton acceleration site is inside these regions, as they are at ultraviolet wavelengths. The resultant $\gamma p$ optical depth for PeV-energy neutrino production is $\gtrsim 10^{-2}$ \cite{Bottcher2009ApJ703p1168,Dermer2012ApJ755p147}. However, several gamma-ray observations have indicated that particle acceleration happens outside broad line regions \cite{Aleksic2011ApJ730L8} (but see \cite{Dermer2012ApJ755p147}). In that case, UHE protons mainly interact with infrared photons of $E_{\gamma} \sim 0.1$ eV from dusty tori and  the neutrinos have higher energies than PeV [Eq.~(\ref{eq:neut_energy_range})], although this photon field is dense enough to produce neutrinos. Hence, as long as protons are accelerated in broad line regions, FSRQs can satisfy the constraint of the $\gamma p$ optical depth from the UHECR flux at 10 PeV (Fig.~\ref{fig:constraints_10PeV}).

FSRQs are also capable of streaming out UHECR protons with enough intensity to yield the detected PeV neutrinos. The luminosity function of AGN is generally described by a broken power-law function with a luminosity-dependent density evolution model \cite{Ueda2003ApJ598p886}. Around the break at which the isotropic equivalent gamma-ray luminosity of $L_{\gamma} \sim 10^{48}$ erg s$^{-1}$, the gamma-ray luminosity function of FSRQs is cosmologically evolved as $\propto (1 + z)^{6.51 \pm 1.97}$ up to $z \sim 1.4$ based on FSRQs detected by {\it Fermi} Large Area Telescope \cite{Ajello2012ApJ751p108} or $\propto (1 + z)^{4.23 \pm 0.39}$ up to $z \sim 1.9$ \cite{Inoue2009ApJ702p523}, where these indices are consistent within the large errors. Since FSRQs with $L_{\gamma} \sim 10^{48}$ erg s$^{-1}$ (called {\it typical} FSRQs below) are energetically dominant, the FSRQ contribution to the total UHECRs can be estimated from this luminosity and the cosmological evolution. With $\psi(z) \propto (1 + z)^4$ up to $z = 2$ and the measured cosmic-ray flux by IceTop, the local UHECR luminosity density is estimated to be $\sim 10^{46}$ erg Mpc$^{-3}$ yr$^{-1}$. This is compatible to the $\gamma$-ray luminosity density of typical FSRQs. Since the gamma-ray luminosity may be comparable with cosmic ray luminosity, FSRQs can provide a major fraction of the UHECR flux at 10 PeV, fulfilling the requirements displayed in Fig.~\ref{fig:constraints_10PeV}.

However, if the spectrum of UHECRs producing PeV-energy neutrinos extends up to EeV energies, only bright FSRQs such as $L_{\gamma} \gtrsim 10^{50}$ erg s$^{-1}$ can satisfy the constraint on the $\gamma p$ optical depth, $\tau_0 \gtrsim 0.2$ shown in Fig.~\ref{fig:constraints_1EeV}. Such bright FSRQs are too rare to energetically reproduce the observed UHECR flux at 1 EeV.

It is unlikely that FSRQs are responsible for both the PeV-energy neutrinos and the highest energy cosmic rays due to their strong cosmological evolution. A typical FSRQ has $\tau_0 \sim 10^{-2}$ if a particle acceleration region is inside its broad line region~\cite{Murase2014arXiv1403.4089}. The allowed parameter space of cosmological evolution for an object with $\tau_0 \sim 10^{-2}$ is shown in the upper right shaded region in Fig.~\ref{fig:gzk_constraint}. However, this region is already ruled out by the UHECR flux at 1 EeV and partially by the nondetection of GZK cosmogenic neutrinos.

\subsubsection{Gamma-ray bursts}

GRBs are another strong candidate of UHECR acceleration sites and therefore high-energy neutrino production sites. Here we introduce four representative sites to produce high-energy neutrinos from literature. 

Internal shocks are the most popular sites to produce high-energy neutrinos \cite{grbint,He2012ApJ752p29}. Particle acceleration at internal shocks is believed to produce prompt emission of GRBs with short time scale variability $\sim 0.01$ s. The internal shocks are generated at the place close to the central engine of GRBs, such as core collapse of massive stars for long duration GRBs or the mergers of compact objects for short duration GRBs, and therefore ultraviolet-to-soft X-ray photons in the rest frame, which are target photons to produce PeV neutrinos, exist densely. The benchmark model of a recent study for long duration GRBs \cite{He2012ApJ752p29} indicates $\tau_0 \sim 0.1$. The internal shocks of low-luminosity GRBs, which are 1 or 2 orders of magnitude dimmer but about 2 orders of magnitude more frequent than high-luminosity or regular GRBs, are also promising sites for the production of PeV-energy neutrinos \cite{llgrbs}. Although their darkness provides fewer target photons, the $\gamma p$ optical depth of $\sim 10^{-2}$ -- $10^{-3}$ can still be achieved, depending on the dissipation radius.

The second candidate site is external shocks, leading to GRB afterglow \cite{Waxman2000ApJ541p707,Murase2007PRD76p123001}. Target photons are the synchrotron radiation of electrons accelerated at the shocks interacting with circumstellar medium. The relatively large spatial scale of the external shocks results in small photon density, and then the optical depth is $\tau_0 \sim 10^{-3}$ for protons to produce PeV-energy neutrinos even considering an early phase \cite{Murase2007PRD76p123001}. Furthermore, magnetic fields lower than those in internal shocks lead to less synchrotron cooling of charged decay products, i.e., charged pions and muons, and therefore the energy spectrum of neutrinos peaks at high energies $\sim 10^9$ -- $10^{10}$ GeV.

The third possible site is inside progenitor stars. Some long duration GRBs with jets of a relatively small Lorentz factor may fail because relativistic jets cannot penetrate progenitor stars for the success of GRBs. Protons can be accelerated in choked jets inside the stars and then produce high-energy neutrinos via photomeson production interactions with dense stellar matter \cite{failedGRBs}. In such dense environments, the efficiency of $p\gamma$ interactions is extremely high, $\tau_0 \gg 1$, which means that all the proton's energy is converted into neutrinos; i.e., the observed UHECRs and neutrinos are not directly connected.

Neutron star binary mergers are believed to be the origin of short GRBs. The mergers also eject significant masses with subrelativistic velocity \cite{NSBMejecta}. The ejecta interact with interstellar medium and produce remnants like supernova remnants \cite{radNSBMRs,Takami2014PRD89063006}. The remnants can accelerate protons up to $\gtrsim 100$ PeV \cite{Takami2014PRD89063006}, and therefore are a potential source of PeV neutrinos. However, target photons for $p\gamma$ interactions, synchrotron radiation of electrons accelerated at the same place, are sparse, and the optical depth for PeV neutrino production is $\tau_0 \sim 10^{-7}$ -- $10^{-5}$.

Among these four possibilities, the internal shock model of regular GRBs and the choked jet model are consistent with the constraint on $\tau_0$ from the UHECR flux at 10 PeV. The internal shock model of low-luminosity GRBs may also be possible. However, their energetics may be problematic. The typical gamma-ray energy output of a regular GRB is $\sim 10^{52}$ erg in gamma rays and the local occurrence rate of long GRBs is $\sim 1$ Gpc$^{-3}$ yr$^{-1}$ \cite{grbrate}, indicating the local CR luminosity of $\sim 10^{44} (\eta_{\rm p} / 10)$ erg Mpc$^{-3}$ yr$^{-1}$ where $\eta_{\rm p}$ is the ratio of UHECR output and gamma-ray output known as the baryon loading factor. The luminosity is about 2 orders of magnitude smaller than that of UHECRs at 10 PeV and thus too low to satisfy the requirements on the UHECR flux for $\tau_0 \gtrsim 10^{-2}$ ($\sim 10$\% of the total cosmic-ray flux; see Fig.~\ref{fig:constraints_10PeV}) unless $\eta_{\rm p} \gtrsim 10^3$. The situation is similar for low-luminosity GRBs because their total energy budget is comparable with that of regular GRBs. Choked jets do not provide UHECRs observed at the earth because of extremely high $\gamma p$ optical depth, and therefore are acceptable if enough such failed GRBs exist.

If the spectrum of UHECRs producing the PeV neutrinos extends to EeV and beyond, the optical depth must be $\tau_0\gtrsim 0.2$ (Fig.~\ref{fig:constraints_1EeV}), leading to the low-luminosity GRBs out of the candidate sources.

It is still valid that GRBs can be the sources of the highest energy cosmic rays with energies of 10 -- 100 EeV. The cosmological evolution of GRBs, although it is still uncertain, is not as strong as FSRQs. In Ref.~\cite{He2012ApJ752p29}, three redshift evolution models are discussed. Two of them are motivated by the cosmic star formation rate and can be approximated by the same functional form as Eq. (\ref{eq:sdf}), but with the power law indices of 2.5 and 3.8 instead of 3.4. The other model mildly evolves to a more distant universe, $\propto (1 + z)^{2.1}$ ($z < 3$) and $\propto (1 + z)^{-1.7}$ ($z \geq 3$). All three of these evolution models are consistent with the constraints by the cosmogenic neutrinos and UHECR intensity shown in Fig.~\ref{fig:gzk_constraint}. Note that the break redshift of the star formation rate evolution can be regarded as our $z_{\rm max}$ as a good approximation in terms of contribution to neutrinos and UHECRs because the star formation rate is no more evolved or negatively evolved above the break redshift. The local UHECR luminosity to reproduce UHECR flux above 10 EeV is $\sim (0.6$ -- $2) \times 10^{44}$ erg Mpc$^{-3}$ yr$^{-1}$ \cite{uhecr19}, which can be marginally provided by regular GRBs. Therefore, regular GRBs can be the origin of the highest energy cosmic rays.

\subsection{\label{sec:mh} Magnetic horizon}

We have constrained the optical depth of $\gamma p$ interactions in the PeV-energy neutrino sources and their cosmological evolution by using a connection between the observed neutrinos and UHECR fluxes. Our estimation of UHECR flux from PeV-energy neutrino sources, i.e., Eq.~(\ref{eq:UHECR_flux}), is valid if the sources are distributed uniformly. However, if UHECR sources are discretely distributed and the Universe is magnetized, low energy cosmic rays cannot reach the Earth within the age of the Universe even from the nearest source, and therefore the UHECR flux at low energies is suppressed without the suppression of neutrino flux~\cite{Lemoine2005PRD71p083007,Aloisio2005ApJ625p249}. This magnetic horizon effect could affect the constraints.

The flux suppression of the arrival UHECRs happens below the energy at which the diffusion length of UHECRs within the energy loss time of cosmic rays $t_{\rm loss}$ is less than the typical separation between the observer and the nearest source $r_{\rm sep}$ \cite{Aloisio2005ApJ625p249}. The diffusion length is $\sim \sqrt{4 D t_{\rm loss}}$ on the assumption of the Kolmogorov diffusion $D(E) = c l_{\rm c} [ r_{\rm L}(E) / l_{\rm c} ]^{1/3} / 3$ where $l_c$ is the coherent length of average extragalactic magnetic fields and $r_{\rm L}(E)$ is the Larmor radius of protons with the energy of $E$ in the extragalactic magnetic field with the strength of $B$. The separation is related to the comoving (local) number density of UHECR sources $n_{\rm s}$.

The number density $n_{\rm s} \sim 10^{-4}$ Mpc$^{-3}$ is motivated by the anisotropy of UHECRs above $6 \times 10^{10}$~GeV~\citep{nsUHECRs}, which is estimated in small-deflection cases, i.e., proton-dominated composition or extremely weak extragalactic magnetic fields, but there is no trivial reason to stick to this number density for the sources of PeV neutrinos. Since the acceleration of cosmic rays up to $\sim 10$ PeV may be much easier than that to $\sim 10^{20}$ eV and many more source candidates are allowed following the Hillas criterion \cite{hillas1984araa22p425}, a larger number density is possible. Note that if many sources contribute to PeV neutrinos but are localized in rare cosmic structures such as clusters of galaxies ($n_{\rm s} \lesssim 10^{-6}$ Mpc$^{-3}$ \cite{ns_clusters}), the effective number density of the sources is determined by the number density of the rare structures.

The properties of extragalactic magnetic fields are poorly known. A reliable upper limit of the properties originates from the Faraday rotation measurements of polarized radio emission from distant objects, $B \sqrt{l_{\rm c}} \lesssim (1~{\rm nG}) \sqrt{1~{\rm Mpc}}$~\cite{Kronberg1994}. The properties of average extragalactic magnetic fields evaluated from recent numerical simulations, e.g., \cite{Ryu2008Sci320p909}, are $B \sqrt{l_{\rm c}} \sim 0.3$ -- $1.0$ nG Mpc$^{1/2}$ \cite{Takami2012ApJ748p9}, which is close to the upper limit obtained by Faraday rotation measurements.

\begin{figure}
\includegraphics[width=0.95\linewidth,clip]{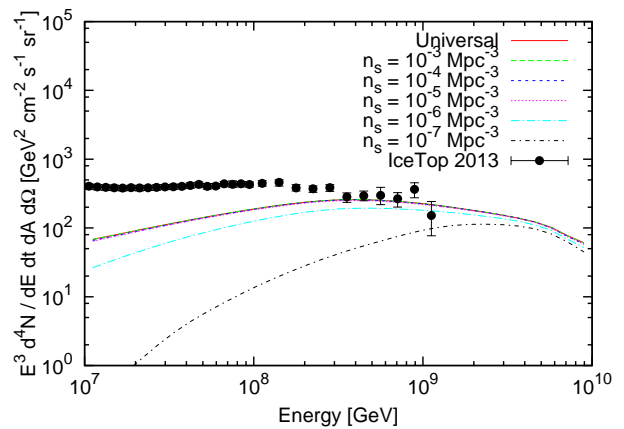}
\caption{UHECR spectra above $10^7$ GeV theoretically calculated with $\alpha = 2.5$, $E_{\rm max} = 10^{10}$ GeV, $B = 1$ nG, and $l_{\rm c} = 1$ Mpc. The universal spectrum, which is calculated from uniformly distributed sources, is normalized to the recent IceTop spectrum \cite{Aartsen2013PRD88p042004} at $10^9$ GeV, corresponding to the local luminosity density of UHECRs with energies between $10^7$ GeV and $10^{10}$ GeV of $\sim 1 \times 10^{46}$ erg Mpc$^{-3}$ yr$^{-1}$. The same local luminosity density is adopted for the other spectra calculated from discretely located sources with the UHECR source number density of $n_{\rm s}$. The magnetic horizon effect becomes significant for rare sources, $n_{\rm s} \lesssim 10^{-6}$ Mpc$^{-3}$ yr$^{-1}$.}
\label{fig:MagHor}
\end{figure}

\begin{figure*}
\includegraphics[width=0.4\linewidth,clip]{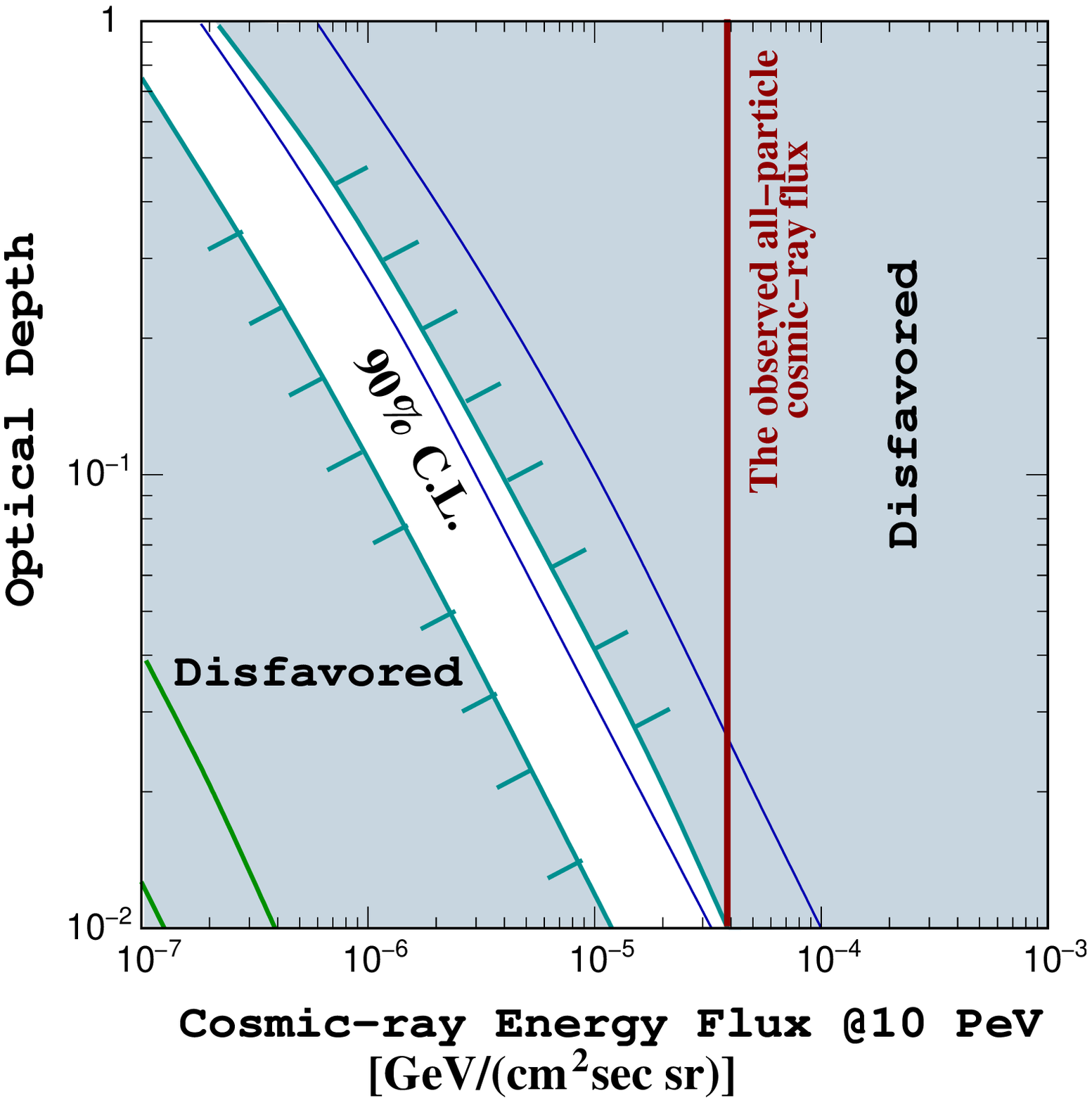}
\includegraphics[width=0.4\linewidth,clip]{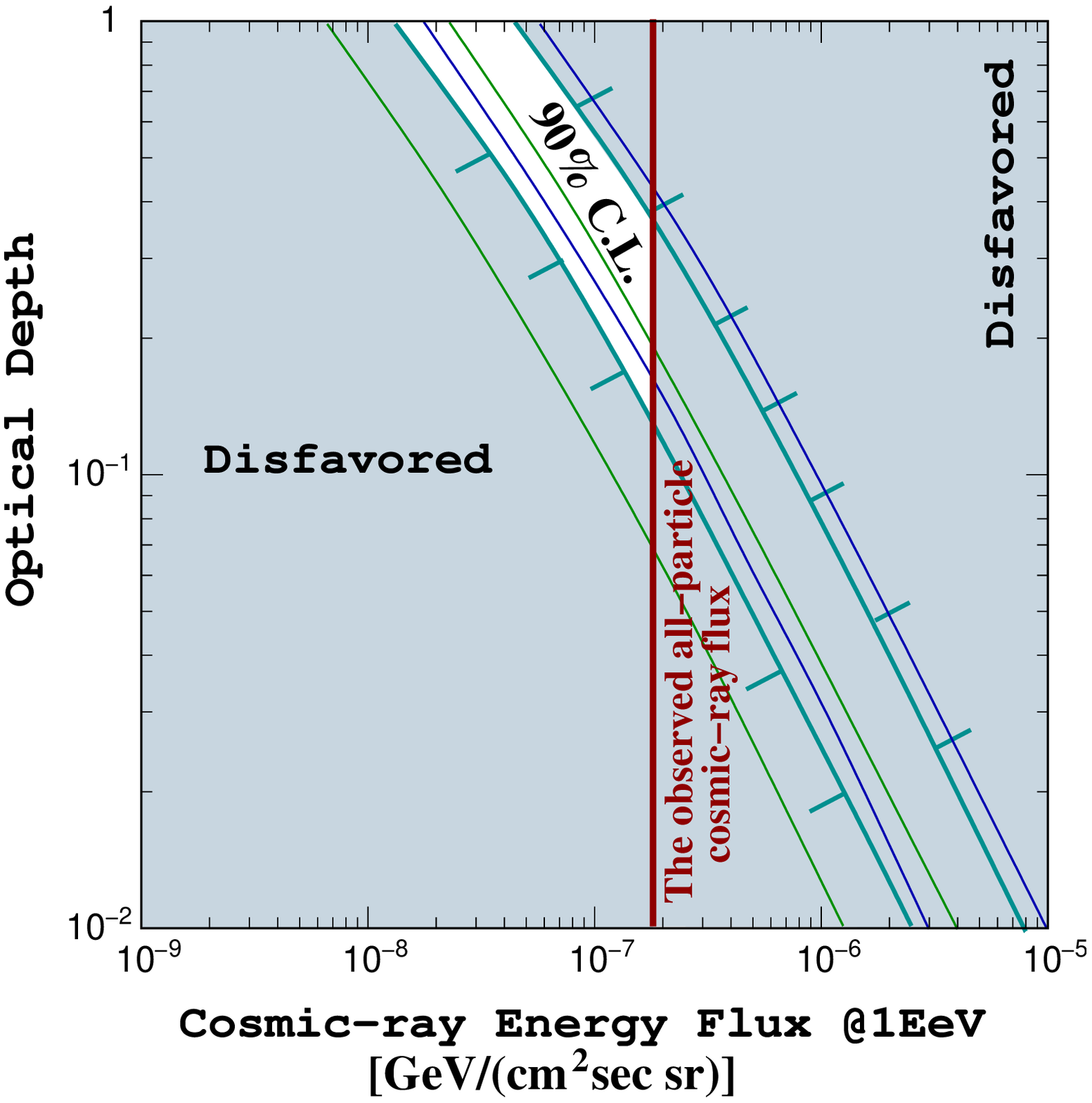}
\caption{Same as Figs. \ref{fig:constraints_10PeV} and \ref{fig:constraints_1EeV}, but the magnetic horizon effect to the constraints is demonstrated. Blue lines show the constraints without extragalactic magnetic fields which are exactly the same as Figs. \ref{fig:constraints_10PeV} and \ref{fig:constraints_1EeV}. The green lines and light-green lines represent constraints including the magnetic horizon effect for $n_{\rm s} = 10^{-6}$ Mpc$^{-3}$ and $n_{\rm s} = 10^{-7}$ Mpc$^{-3}$, respectively. The cases of $n_{\rm s} \geq 10^{-5}$ Mpc$^{-3}$ are the same as the blue lines because the magnetic horizon effect is quite small. Note that the shades are the parameter space disfavored by the neutrino detection and the total cosmic-ray flux for $n_{\rm s} = 10^{-6}$ Mpc$^{-3}$.}
\label{fig:constraints_mh}
\end{figure*}

Figure \ref{fig:MagHor} demonstrates the magnetic horizon effect in UHECR spectra above $10^7$ GeV with the recent IceTop spectrum \cite{Aartsen2013PRD88p042004} for normalization. The theoretical spectra are calculated following a formulation in Ref. \cite{BGdiffusion} with $B = 1$ nG and $l_{\rm c} = 1$ Mpc, but their comoving source term is extended with the cosmological evolution of UHECR sources, $Q(E,z) \propto \psi(z) E^{-\alpha}$. The cosmological evolution is implemented as pure luminosity evolution without changing the comoving density of UHECR sources. The universal spectrum, i.e., the spectrum of UHECRs from uniformly distributed sources, corresponds to Eq.~(\ref{eq:UHECR_flux}). Sources are located at the three-dimensional lattice points $(x, y, z) = (D_{\rm sep} ( i + 1/2 ),  D_{\rm sep} ( j + 1/2 ),  D_{\rm sep} ( k + 1/2 ) )$ for integers $i, j, k$ for discrete sources with the observer at the origin of the coordinates, where $D_{\rm sep} = ( 3 / 4\pi n_{\rm s} )^{1/3}$. The distance to the nearest source(s) is actually $r_{\rm sep} = \sqrt{3} D_{\rm sep} / 2 = 13.4 ( n_{\rm s} / 10^{-4}~{\rm Mpc}^{-3} )^{-1/3}~{\rm Mpc}$. The comoving luminosity density of UHECRs with energies between $10^7$ GeV and $10^{10}$ GeV is $\mathcal{L}_0 = 1 \times 10^{46}$ erg Mpc$^{-3}$ yr$^{-1}$ in the local universe to suit the universal spectrum to the IceTop spectrum at $10^{9}$~GeV. The input spectrum has the index of $\alpha = 2.5$ and is truncated at $10^{10}$~GeV.

The magnetic horizon effect, which appears in the difference between the universal spectrum and a spectrum calculated from discretely distributed sources, is not significant above $10$ PeV for $n_{\rm s} = 10^{-3}$ -- $10^{-5}$ Mpc$^{-3}$ (the propagation theorem \cite{Aloisio2004ApJ612p900}). In these cases, the constraints on $\gamma p$ optical depth, i.e., Figs. \ref{fig:constraints_10PeV} and \ref{fig:constraints_1EeV}, do not change.

If $n_{\rm s}$ is very small, the magnetic horizon effect appears. In the case of $n_{\rm s} = 10^{-6}$ Mpc$^{-3}$, the UHECR flux at $10$ PeV is suppressed by a factor of 2.5 compared to the universal spectrum under the same cosmic ray luminosity density, and therefore, the constraint on $\tau_0$ (Fig. \ref{fig:constraints_10PeV}) is relaxed. The UHECR flux suppression is found also at 1 EeV, but by 20\%. Thus, the constraint from UHECR flux at 1 EeV (Fig. \ref{fig:constraints_1EeV}) is nearly unchanged. These magnetic horizon effects to the constraints on $\tau_0$ are summarized in Fig.~\ref{fig:constraints_mh}.

In the case of $n_{\rm s} = 10^{-7}$ Mpc$^{-3}$, the magnetic horizon effect becomes much larger. The UHECR fluxes at $10$ PeV and 1 EeV are suppressed by a factor of 300 and 3 compared to the universal spectrum, respectively (see Fig. \ref{fig:constraints_mh} again). These flux suppressions relax the constraints, and therefore a much smaller $\gamma p$ optical depth can be allowed as long as source candidates can provide UHECRs enough to produce PeV-energy neutrinos.

The flux suppression mainly depends on the properties of extragalactic magnetic fields. Thus, suppression factors for $\alpha = 2.3$ and $2.7$ are close to those discussed for $\alpha = 2.5$.

The magnetic horizon effect can affect some results discussed in the previous subsection. For example, LBLs can satisfy a constraint on $\tau_0$ at the cosmic-ray energy of 10 PeV. A recent gamma-ray population study revealed that the cosmological evolution of LBLs is weakly positive in number density, but can be well modeled by pure luminosity evolution with the local number density of $(3$ -- $5) \times 10^{-7}$ Mpc$^{-3}$ \cite{Ajello2014ApJ780p73}. In this case, the allowed region in the left panel of Fig. \ref{fig:constraints_mh} is between the green region and the light-green region, and the estimated optical depth, $\tau_0 \sim 3 \times 10^{-4}$ becomes consistent with the constraint thanks to the magnetic horizon effect. However, LBLs are not energetically enough to contribute to a dominant fraction of extragalactic cosmic rays at 10 PeV, which is also required from the left panel of Fig. \ref{fig:constraints_mh}.

\subsection{$pp$ interactions}

We have examined neutrinos by $\gamma p$ interactions, but $pp$ interactions also can produce neutrinos. The hadronuclear interactions become important in the systems where there is no dense photon fields, but UHECRs are in abundance and efficiently confined, such as starburst galaxies \cite{sbn,Murase2013PRD88p121301} and clusters of galaxies \cite{clustern}. Remarkable differences of $pp$ interactions from $\gamma p$ interactions are a much lower energy threshold and no prominent resonant structure in the cross section of $pp$ interactions. As a result, nonthermal protons distributed over a wide energy range produce secondary neutrinos and gamma rays also over a wide energy range. Reference \cite{Murase2013PRD88p121301} showed that the secondary gamma-ray emission can violate the flux of diffuse gamma-ray emission detected by the {\it Fermi} Large Area Telescope unless the spectrum of cosmic-ray protons is rather hard. Also, such a hard spectrum may easily violate the indicated IceCube cutoff at several PeV energies in the near future \cite{icecubeHESE2013}. Thus, $pp$ sources will be strongly constrained in future multimessenger studies.

\subsection{Mass composition and source individuality}

Although we have constrained the properties of UHECR and neutrino sources focusing only on protons, it is possible that a significant fraction of heavier nuclei are involved in the observed cosmic rays. Mass composition of cosmic rays could affect some of the derived bounds.

The bounds of $\gamma p$ opacity and extragalactic cosmic-ray proton intensity are independent of cosmic-ray mass composition. The $\gamma p$ optical depth of UHECR sources are constrained by the total observed intensity of UHECRs (including both protons and heavier nuclei), which is shown as the red vertical lines in Figs. \ref{fig:constraints_10PeV}, \ref{fig:constraints_1EeV}, and \ref{fig:constraints_mh}. If the observed UHECRs contain a significant fraction of heavier nuclei, cosmic-ray proton intensity to be adopted for the constraints becomes smaller than that shown in the red lines, and then the allowed parameter region becomes smaller. Thus, the bounds that have been derived in Sec. \ref{subsec:optdepth} are the conservative limits of $\gamma p$ opacity of UHECR sources and extragalactic cosmic-ray proton intensity.

On the other hand, the constraints on the cosmological evolution of UHECR sources (Fig. \ref{fig:gzk_constraint}) rely on the mass composition of the highest energy cosmic rays because the intensity of cosmic-ray protons is normalized with the observed UHECR intensity at 100 EeV. A sizable fraction of heavy-nucleus cosmic rays suggested by the Pierre Auger collaboration in the highest energy range \cite{AugerHeavy} relaxes our constraints, although the mass composition of the highest energy cosmic rays is still a big mystery as shown in the uncertainty of hadronic interaction models (e.g., \cite{Allen2013ICRC1182}), the anisotropy measurements at low energies \cite{PAO2011JCAP06p022}, and the interpretation of the data of the Telescope Array experiment \cite{TAcomposition}. If heavy nuclei are included in the highest energy cosmic rays, cosmic-ray proton intensity is reduced and therefore the stronger cosmological evolution of UHECR sources, i.e., larger $m$ and $z_{\rm max}$ become allowed. An accurate measurement of the UHECR proton flux will improve our knowledge of the source evolution of UHECRs obtained by the cosmic neutrino detection.

The observed cosmic-ray spectrum integrating the injection spectrum over all the sources can generally have a different spectral slope from that of an individual injection spectrum. On the assumption of identical sources, this spectral modification is properly considered in Eq. (\ref{eq:UHECR_flux}), which takes adiabatic energy-loss due to cosmic expansion. This is a dominant energy-loss process below 1 EeV. Note that all the possible energy-loss processes are considered by solving a diffusion equation numerically when the magnetic horizon effect is discussed (Sec.~\ref{sec:mh}). However, modification by other factors, such as cosmic-ray escape from magnetized structures in the sources and the possibility that individual sources have different maximum injection energy of cosmic rays, is not included. Nevertheless, the bounds of the $\gamma p$ optical depth and extragalactic cosmic-ray proton intensity derived from the total cosmic-ray intensity at 10 PeV are rather solid because the neutrinos detected by IceCube are produced by cosmic-ray protons with energies around 10 PeV.

Even in the cases that allowed $\tau_0$ to be evaluated with cosmic-ray intensity at 1 EeV and that the cosmological evolution of UHECR sources has constrained with cosmic-ray intensity at $10^{20}$ eV, our bounds are approximately valid by considering an effective power-law index. A resultant total cosmic-ray spectrum modified by propagation effects and/or individual source properties may have some curvature, but this can be approximated by a power-law function with an appropriate spectral index for a limited energy range in question. For example, when the $\gamma p$ optical depth is constrained by the observed cosmic-ray intensity at 1 EeV, it is enough to consider a power-law function intersecting the calculated cosmic-ray spectrum at 10 PeV and 1 EeV. Typically, we can regard the dependence of the present opacity bounds on the spectral index $\alpha$ ranging from 2.3 to 2.7 as the variance of the present constraints due to possible departure from the global single power-law spectrum assumption.

Especially, in the cases where the maximum energy of protons is different among sources and distributes following a power-law function, i.e., $dn / dE_{\rm CR,max} \propto E_{\rm CR,max}^{-\beta}$, down to 10 PeV, our formulation correctly derives the opacity bounds. A power-law spectrum of cosmic rays with the index of $\alpha$ at generation and a power-law distribution of the maximum energy of cosmic rays result in a power-law spectrum of total cosmic rays with the index of $\alpha + \beta - 1$ \cite{Kachelriess2006PLB634p143}. The derived constraints can be interpreted by considering identical sources with this index effectively.

\section{\label{sec:summary} Summary}

We have developed an analytical formula for the intensity of PeV-energy neutrinos, and have derived the general constraints on the optical depth of $\gamma p$ interactions in the neutrino sources, required fractions of extragalactic cosmic rays, and on their cosmological evolution, assuming that the photomeson production of extragalactic cosmic rays in the sources provides the bulk of PeV-energy neutrinos detected by IceCube. The connection between the IceCube neutrino flux and cosmic-ray flux at $10$ PeV, which corresponds to the energy of protons producing the IceCube neutrinos, indicates that the $\gamma p$ optical depth is $\gtrsim 10^{-2}$ and that extragalactic component of UHECRs contributes to more than a few percent of cosmic ray flux at 10 PeV. These constraints become more stringent if the PeV-energy neutrino sources are also responsible for EeV-energy cosmic rays, favoring the dip transition model of cosmic rays. Furthermore, if these sources are also the emitters of the highest energy cosmic rays with energies reaching to 100 EeV, they should not be strongly evolved with cosmic time. Among the source candidates of the PeV-energy neutrinos discussed in this paper, FSRQs satisfy the conditions as sub-PeV to PeV energy cosmic ray and neutrino emitters, but they cannot simultaneously be emitters of the highest energy cosmic rays due to their strong cosmological evolution. The internal shocks at regular GRBs can account for the neutrino detection only if the rather extremely large baryon loading takes place in the GRB emission. None of the known extragalactic objects can be found to function as an origin of both PeV-energy neutrinos and the highest energy cosmic rays. The obtained constraints are valid even in the magnetized universe unless the UHECR sources are very rare in space, such as their number density is smaller than $10^{-6}$ Mpc$^{-3}$.

\begin{acknowledgments}
We wish to acknowledge M.~Ahlers, K.~Ioka, and K.~Murase for useful discussions in preparations for this work. We also thank M.~Relich for helpful comments on the manuscript and M.~Bustamante for useful comments. This work is supported by the JSPS Grants-in-Aid for Scientific Research No. 25105005, No. 25220706 (S.~Y.), and No. 24$\cdot$9375 (H.~T.). 
\end{acknowledgments}

\bigskip

\appendix

\section{Integration on source redshift} \label{sec:app1}

We approximately integrated the neutrino yield over redshift in Sec. \ref{sec:function}. The treatment of the integration follows that in Ref. \cite{uhecr_constraint2012}. If $\Omega_{\rm M} (1 + z)^3 \gg 1$, which is valid in the wide range treated in this study, we can derive the following expression by using $t \equiv \sqrt{\Omega_{\rm M} (1 + z)^3 + \Omega_{\Lambda}}$: 
\begin{widetext}
\begin{equation}
I_x \equiv \int \frac{(1 + z)^x dz}{\sqrt{\Omega_{\rm M} (1 + z)^3 + \Omega_{\Lambda}}} \simeq \frac{2}{2 x - 1} \Omega_{\rm M}^{- \frac{x + 1}{3}} 
\left\{ \Omega_{\rm M} (1 + z)^3 + \Omega_{\Lambda} \right\}^{\frac{x}{3} - \frac{1}{6}}. 
\label{eq:app1}
\end{equation} 
Also, a useful application for this study is 
\begin{equation}
\int \frac{(1 + z)^x \log (1 + z) dz}{\sqrt{\Omega_{\rm M} (1 + z)^3 + \Omega_{\Lambda}}} \simeq \frac{2}{2 x - 1} \Omega_{\rm M}^{- \frac{x + 1}{3}} 
\left\{ \Omega_{\rm M} (1 + z)^3 + \Omega_{\Lambda} \right\}^{\frac{x}{3} - \frac{1}{6}} \log (1 + z) - \frac{2}{2 x - 1} I_x, 
\end{equation} 
if the derivative of Eq. (\ref{eq:app1}) is applied. 
\end{widetext}


\end{document}